\documentclass[twocolumn,amsmath,amssymb,superscriptaddress]{revtex4-1}
\usepackage[utf8]{inputenc} 

\usepackage{geometry} 
\usepackage{graphicx} 

\usepackage{booktabs} 
\usepackage{array} 
\usepackage{paralist} 
\usepackage{verbatim} 
\usepackage{subfig} 
\usepackage{epsf}
\usepackage[colorlinks=true,linkcolor=blue,urlcolor=blue,citecolor=blue,pdfusetitle]{hyperref}
\hypersetup{colorlinks,linkcolor={blue},citecolor={blue},urlcolor={blue}} 

\usepackage{fancyhdr} 
\usepackage[nottoc,notlof,notlot]{tocbibind} 
\usepackage[titles,subfigure]{tocloft} 

\newcommand{\bla}{bla\\bla\\bla\\bla\\bla}

\begin{document}
\title{Assessment of Atomic and Molecular Physics in Africa}
\author{St\'ephane Kenmoe}
\affiliation{Department of Theoretical Chemistry, University of Duisburg-Essen,\\ Universit\"atsstr. 2, Essen D-45141, Germany}.
\author{Obinna Abah}
\affiliation{School of Mathematics, Statistics and Physics, Newcastle University, Newcastle upon Tyne NE1 7RU, United Kingdom}.

\begin{abstract}
We present the status of the research in the field of atomic and molecular physics in Africa as well as some challenges hindering the efforts being made by the African scientists. We further report the discussions and progress of the  African Strategy for Fundamental and Applied Physics (ASFAP) working group on Atomic and Molecular physics with the view of providing the continent research direction for next decade.
\end{abstract}

\maketitle

Recent advances in experimental and theoretical scanning probing methods at the atomic scale have led to tremendous applications in biology, medicine, electronics, quantum technologies, spintronics or heterogeneous catalysis. For example, insight into the structure of living cells, the single molecule transistor, the minute working of catalytic reactions allowing the rational design of catalysts and improvement of properties, just to cite a few. However, probing matter at the nanoscale on the African continent is still challenging, both theoretically and experimentally. This stems from the various limitations in research facilities. 
 
\begin{figure}[!h]
\begin{center}
(a)
\includegraphics[width=0.5\textwidth]{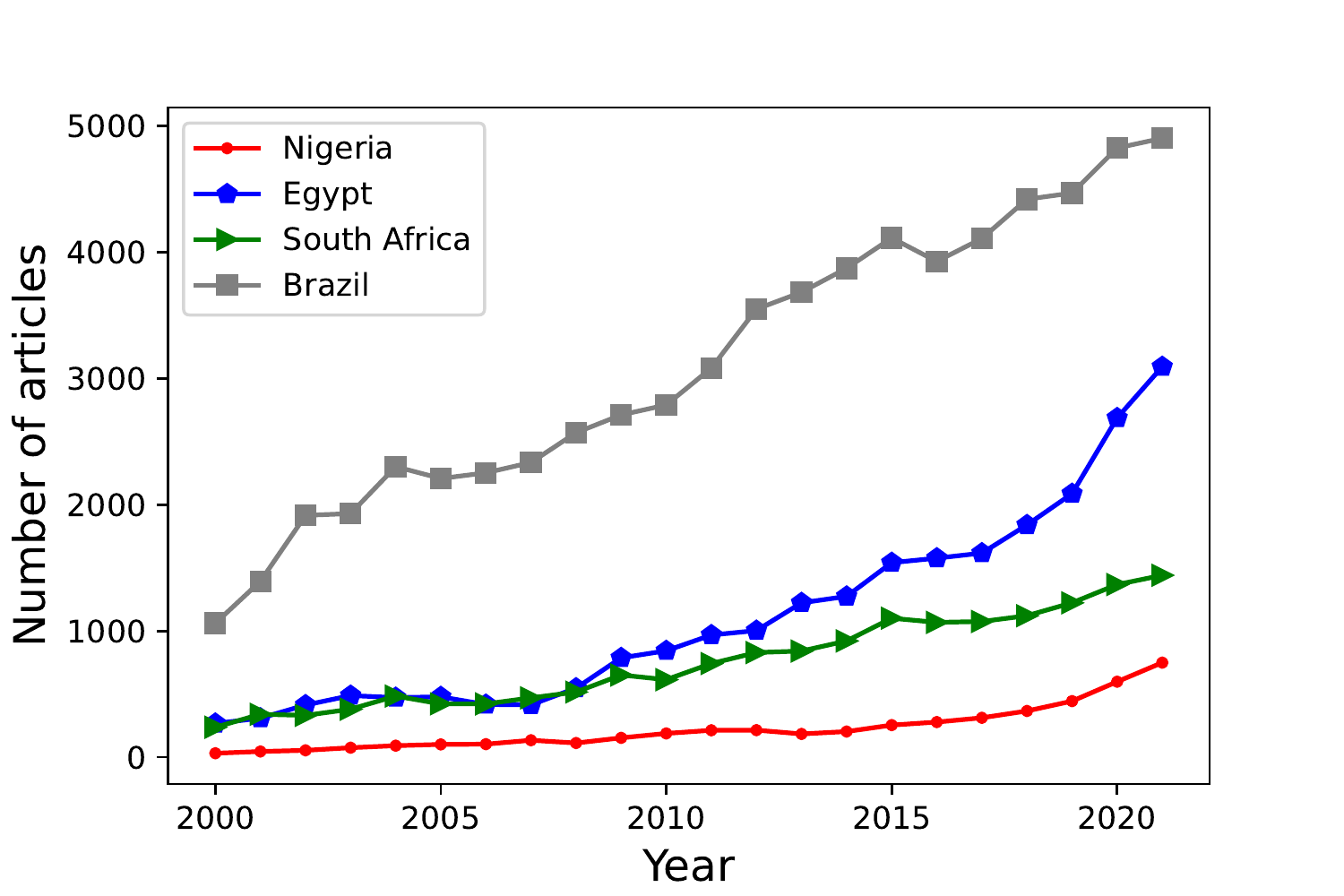}
(b)
\includegraphics[width=0.5\textwidth]{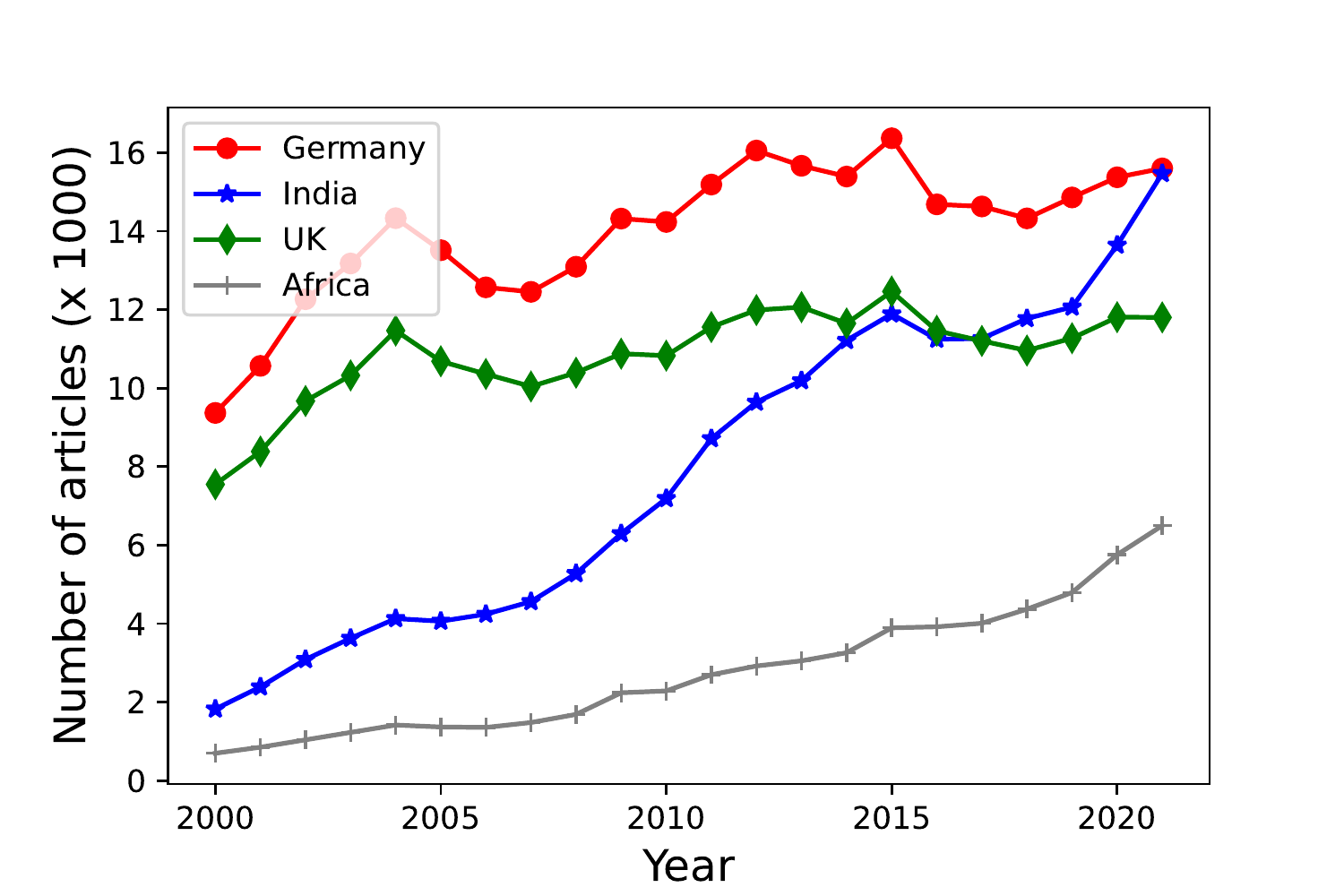}
\end{center}
\caption{\label{figure} Research output per year from 2000 – 2021 for search keywords: atoms, atomic, molecular, molecules, or ions. (a) The number of articles published by some African countries (Egypt, Nigeria, South Africa) compared to the Brazil.  (b) The total articles published by African scientists (Algeria, Cameroon, Congo, Egypt, Ethiopia, Ghana, Kenya, Morocco, Nigeria, South Africa, Tunisia) compared western countries (Germany and UK) and India. Source: Scopus -- accessed October 8, 2022.\cite{scopus} }
\end{figure}

Despite the population of about 1.3 billion, which are mainly youth, the research and development output of Africa is quite low in virtually all areas of physics. To quantitatively understand this abysmal performance, we analyse the amount of research articles published by African scientists (based in African institutions) from 2000 – 2021, see Figure \ref{figure}. Over the last two decades, the total research output from Africa stands shy of 70,000 articles with about 6,000 per year in recent times. It will be interesting to know that these are comparable to the Brazil scientific research output over the same period. However, the dramatic rise of India over the same period clearly shows the need for understanding the problem facing African scientists. This graphical illustration could readily be linked to the poor economic performance of the Africa continent, the world’s poorest inhabited continent according to the World Bank. This is basically demonstrated by the difficulty to access energy for community services (health, education and so on) as well as the lack/inadequate information and communication technologies among others \cite{nature2019}. Moreover, only Egypt and South Africa made it in the Top 40 of the world's research and development index in 2021 \cite{RD2021}.  However, Africa Union Agenda 2063 has identified Physics – fundamental and applied as a key solution to address the developmental problems facing the continent \cite{AU2063}.
\subsection*{Challenges facing African scientists/physicists} 
On a theoretical point of view, electrical power instability in many countries does not allow sustainable computing and computational facilities are scarce, see Ref.\cite{Mulilo2022} for more discussion. Most sub-saharan countries barely have supercomputers available for research. The few available facilities on the continent are concentrated in Northern Africa and South Africa. Researchers rely on the latter and on external partners such as the Abdus Salam International Centre for Theoretical Physics, Italy. A dependence that limits the productivity but also the size of the system to study simple molecules.  Experimentally, resources are also scarce. For example, it is only recently that central Africa got its first operational AFM apparatus in what is likely the first nanotechnology laboratory in the Republic of Congo. Besides, the light source community is still to build the first synchrotron on the continent and relies on external sources and networks like the  Synchrotron-Light for Experimental Science and Applications in the Middle East (SESAME) and the free and open-source software such as Large-scale Atomic/Molecular Massively Parallel Simulator (LAAMPS). Unfortunately, for Africa, international organizations often support research of their interest and are compounded by the government's ill-advised policies towards education.

\subsection*{Current support towards enhance research output} 
During the last decades, various research groups and networks have been active on the continent, thanks to some foreign collaborations/donors. These include Physics Department, Marien Ngouabi University (Brazzaville, Congo), CEPAMOQ (Douala, Cameroon), Lasers Atoms Laboratory, Cheikh Anta Diop University (Dakar, Senegal), Atomic Molecular Spectroscopy and Applications Laboratory, University of Tunis El Manar (Tunisia), Medical University of Southern Africa (South Africa), African Laser Atomic Molecular and Optical Science Network. In addition, there is growth in the study of materials sciences in Africa through the African School for Electronic Structure Methods and Applications (ASESMA).

As an extension of these efforts, African physicists from a variety of specializations are developing an African strategy for basic and applied physics, see https://africanphysicsstrategy.org/ \cite{ASFAP}.  Organized into several working groups, committees, and forums, they are working to produce a report to inform the African and broader community of strategic directions that can positively impact physics education and research over the next decade \cite{ketevi2022,fassi2022}.The report is intended to help African policy makers, educators, researchers, communities, and international partners prioritize resources and activities for physics education and research at the national, regional, and pan-African levels. As part of this group of African physicists, we have the task of coordinating the activities of the Atomic and Molecular Physics working group.

\subsection*{Atomic and molecular physics working group – journey so far and way forward}
In the spirit of the ASFAP, the Atomic and Molecular Physics (AMP) working group aims at reporting on the state of research and knowledge transfer of these groups and their derivatives on the continental level but also on the various research carried by African scientists in AMP performed all over the world and that align to sustainable development goals. From the above-mentioned research groups and networks, we have identified and have traced the various African scientists still active in the field, their research interests and compiled their various achievements.

As part of this, we have successfully organised meetings and had an online workshop on Atomic and Molecular Physics in January 2022 during which the discussion is cantered on identifying challenges facing different research groups across the continent among others. These efforts, in conjunction with other ASFAP working group, have resulted in some letter of intents (LOIs) submitted for the strategies. In addition, after deliberation with the ASFAP Steering committee members and the Photonics and Optics working group during the second African Conference of Fundamental and Applied Physics ACP2021, there is a unilateral decision to merge the two working groups – \textit{Atomic, Molecular and Optical Physics}. We believe that this will synergise interdisciplinary activities towards industrial and technological advancements.

To conclude, we advocate for physics-based policies in the various country, region and the continent at large. These will be geared towards development of human capital as well as engaging the private sectors for support. Finally, with the support of international collaborations, qualitative increase in the research output of Atomic, Molecular and Optical Physics in Africa will become a fruition.


\renewcommand{\bibname}{References}

\end{document}